\documentclass[draftclsnofoot,onecolumn,12pt,oside]{IEEEtran}
\usepackage{amssymb}
\usepackage{cite}
\usepackage{xcolor}

\ifCLASSINFOpdf
   \usepackage[pdftex]{graphicx}
\else
   \usepackage[dvips]{graphicx}
\fi
\usepackage{color,soul}
\usepackage{amssymb}

\usepackage{amsmath}

\usepackage{epstopdf}

\begin{document}
%
\title{A Joint Code-Frequency Index Modulation for Low-complexity, High Spectral and Energy Efficiency Communications}
%
%
%

\author{Minh~Au,~\IEEEmembership{Member,~IEEE,}      
        Georges~Kaddoum,~\IEEEmembership{Member,~IEEE,}
        Francois~Gagnon,~\IEEEmembership{Senior~Member,~IEEE}
        and~Ebrahim~Soujeri,~\IEEEmembership{Senior Member,~IEEE,}
\thanks{M. Au, G. Kaddoum, F. Gagnon and E. Soujeri are with the Department
of Electrical Engineering, \'Ecole de Technologie Sup\'erieure, Montr\'eal, QC, H3C 1K3 CA e-mail: minh.au@lacime.etsmtl.ca.}
\thanks{Manuscript received November 20, 2017; revised December 26, 2017.}}

%
%

\markboth{IEEE Transactions on Communications}%
{Shell \MakeLowercase{\textit{et al.}}: Bare Demo of IEEEtran.cls for IEEE Journals}
%



\maketitle

\begin{abstract}
A relatively simple low complexity multiuser communication system based on simultaneous code and frequency index modulation (CFIM) is proposed in this paper. The proposed architecture reduces the emitted energy at the transmitter as well as the peak-to-average-power ratio (PAPR) of orthogonal frequency-division multiplexing (OFDM)- based schemes functions without relegating data rate. In the scheme we introduce here, we implement a joint code- frequency- index modulation (CFIM) in order to enhance spectral and energy efficiencies. After introducing and analysing the structure with respect to latter metrics, we derive closed-form expressions of the bit error rate (BER) performance over Rayleigh fading channels and we validate the outcome by simulation results. Simulation are used verify the analyses and show that, in terms of BER, the proposed CFIM outperforms the existing index modulation (IM) based systems such as spatial modulation (SM), OFDM-IM and OFDM schemes significantly. To better exhibit the particularities of the proposed scheme, PAPR, complexity, spectral efficiency (SE) and energy efficiency (EE) are thoroughly examined. Results indicate a high SE while ensuring an elevated reliability compared to the aforementioned systems. In addition, the concept is extended to synchronous multiuser communication networks, where full functionality is obtained. With the characteristics demonstrated in this work, the proposed system would constitute an exceptional nominee for Internet of Things (IoT) applications where low-complexity, low-power consumption and high data rate are paramount. 
\end{abstract}


%
\IEEEpeerreviewmaketitle

\section{Introduction}
\IEEEPARstart{F}{acing} a tremendous demand for higher data rate communication systems associated with an increasing need in the number of devices and gadgets for mobile Internet and Internet of Things (IoT) applications, is the challenge that needs to be addressed in the next generation of wireless systems. Indeed, they are required to achieve high spectral efficiency (SE) while maintaining massive connectivity. For some emerging devices such as drones and tiny sensors in intelligent vehicle technologies, ultra reliable and extremely low-latency are absolute requirements. 

Within the development of future wireless networks, the concept of IoT has emerged as the most promising technology that supports and enables many devices to gain Internet connectivity. This emerging technology has only become available via major advances in and availability of tiny, cheap and intelligent sensors that are cost-effective and easily deployable \cite{Rawat2014}. Guaranteeing battery life saving by having a low energy consumption should be another noteworthy feature of these IoT devices. A large number of researches have devoted themselves to fulfil these requirements over the last decade \cite{Tozlu2012,Chen2012,Toh2001}. 

Recently, the concept of index modulation (IM), which utilizes the indices of some transmission entities to carry extra information bits, has been proposed as a competitive alternative in digital modulation techniques for the future wireless communications \cite{Basar2016}. These IM-based systems aim to increase data rate and energy efficiency (EE) in emerging wireless communication devices. In this direction, research in the last decade has seriously considered the usage of certain parameters such as space, polarity, code and frequency as modulation indices to enable the transmission of extra bits per symbol without demanding larger bandwidth or higher power.

In this vein, the spatial modulation (SM) technique which is based on transitioning between multiple antennas and using the index of the active antenna as a measure to carry extra information is an area of growing interest where fruitful results have been provided. \cite{Mesleh2006,Jeyadeepan2008,Mesleh2008,Renzo2014,Yang2015}. However, this technique requires additional antennae to achieve more data rate. So, it is not implementable in sensor networks. 
Other examples of index-based modulation systems that involve orthogonal frequency-division multiplexing (OFDM) include subcarrier-index modulation (SIM)-OFDM \cite{Abualhiga2009} and enhanced subcarrier-index modulation (ESIM)-OFDM \cite{Tsonev2011}. However these systems are primitive and stay short of conventional OFDM systems in terms of overall performance.

A relatively stronger but more complex index-based modulation system that involves OFDM is called OFDM with Index Modulation (OFDM-IM) and is proposed in \cite{Basar2013}. In this approach, the subcarriers are grouped into many blocks where a subdivision of subcarriers contained in each block is \emph{turned on} depending on indexing bits, where the activated subcarriers transmit a symbol each. This OFDM-based IM technique utilises maximum likelihood (ML) detection at the receiver which is indeed too complex for IoT and wireless sensor devices because low complexity and low power consumption say the final word in these tiny gadgets.

As the IM modulation schemes mentioned above are either hardware-demanding or too complex, very low-complexity and energy efficient IM techniques have been proposed more recently \cite{Kaddoum2016,Kaddoum2015a,Kaddoum2015,Soujeri2017}. In the generalized code index modulation (GCIM) \cite{Kaddoum2016,Kaddoum2015a,Kaddoum2015}, extra information bits are carried by the selection of spreading codes. On the other hand, frequency index modulation (FIM) \cite{Soujeri2017} is an OFDM system based, in this scheme extra bits are carried by the activation of one subcarrier, while the rest are nulled out. By choosing orthogonal spreading codes and frequencies, these bits are estimated based on square law energy detection (SLED) at the receiver side, yielding low-complexity and high EE. However, neither systems achieve a high SE because GCIM is designed for single-carrier communications, while some subcarriers are not used in FIM.

Furthermore, in terms of multiuser communication perspective, given that 5G standards promise massive connectivity for billions of devices \cite{Boccardi2014,Dai2015}, these aforementioned systems might be limited and cost-ineffective because SM requires hardware, i.e many antennae, while more bandwidth is needed in OFDM-based systems incorporating IM. Therefore, achieving high SE in FIM and GCIM is imperative. To the best of our knowledge, there is no performance analysis of IM systems for multiuser communications. Moreover, no attempt has been made regarding the benefit of having several transmission entities to carry extra information bits. 

Motivated by the aforementioned problems, this paper exploits the benefits of having a joint code-frequency IM scheme that could meet the needs of 5G wireless systems to a great extent. This paper extends the idea of GCIM to multicarrier systems to enhance SE while maximizing EE by reducing the peak-to-average power ratio (PAPR) via FIM. Both of these schemes exhibit low-complexity and can be used for multiuser communications via spreading codes. The combination of these suit IoT applications where several sensors and devices must be connected. 

The contributions of this paper are summarized as follows:
\begin{itemize}
\item The joint frequency and spreading code indexing is used as an information-bearing unit to increase both SE and EE without adding extra hardware complexity to the system.
\item A low-complexity detection method based on SLED is proposed.
\item Closed-form mathematical expressions for the probability of bit error of CFIM are derived. The performance of CFIM is compared to other IM-based approaches such as SM, OFDM-IM as well as conventional methods such as OFDM. Results obtained show that CFIM outperforms these schemes. 
\item Under the framework of multicarrier systems, CFIM scheme is analysed in synchronous transmission for the first time. 
\end{itemize}

The remainder of this paper is organized as follows. Section \ref{sect:II} provides the system model of CFIM including the transmitter and receiver structures. Closed form expressions of the probability of bit error are analyzed in Section \ref{sect:III}. SE, EE and complexity of the system are discussed in Section \ref{sect:IV}. CFIM is extended to synchronous multiuser communications in Section \ref{sect:V}. Section \ref{sect:VI} presents computer simulation results concerning SE, EE and complexity in comparison to SM, OFDM-IM and conventional OFDM systems. Conclusions follow in Section \ref{sect:VII}.

\section{ System Model}
\label{sect:II}
In this section, we present the CFIM transmitter and receiver architectures.

\subsection{The Transmitter}
\label{sect:II.a}
\begin{figure*}[!t]
\centering
\includegraphics[width=7.0in,height=3.0in]{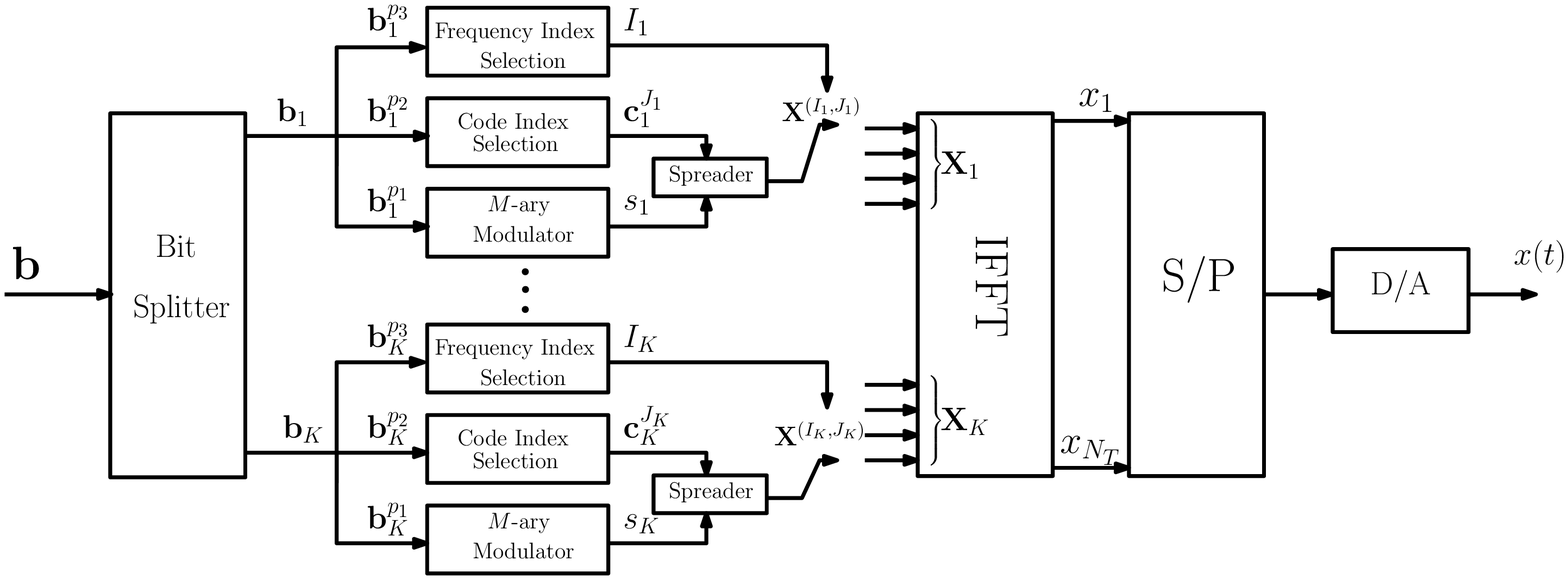}
\caption{CFIM architecture of the transmitter}
\label{fig:1.1.CFIM_Tx}
\end{figure*}

A typical architecture of the CFIM transmitter with $K \times N$ subcarriers is depicted in Fig. \ref{fig:1.1.CFIM_Tx}. A sequence $\mathbf{b}$ of $p_{T}$ bits is equally separated into $K$ blocks. Since each CFIM block has the same processing procedure, we consider the $k^{\rm th}$ block for simplicity hereafter. In this block, the sequence $\mathbf{b}_{k} = \left[\mathbf{b}_{k}^{p_{1}},\mathbf{b}_{k}^{p_{2}},\mathbf{b}_{k}^{p_{3}} \right] $ leaves the transmitter in chunks or sub-blocks of $p$ bits where each sub-block consists of 3 sub-chunks such that $p = p_{1}+p_{2}+p_{3}$. 
In this format, $\log_{2}(M) = p_{1}$ denotes the number of modulated bits that map into an $M-$ary signal constellation to produce a symbol $s_{k} \in \mathcal{S}$, where $\mathcal{S}$ is a set of $M-$ary signals. Further, $\log_{2}(N_{c}) = p_{2}$ is the number of mapped bits required to select the spreading code among $N_{c}$ codes, and $\log_{2}(N) = p_{3}$ is the number of mapped bits needed to activate one subcarrier.
In this paper, we consider MPSK modulation only for its higher energy efficiency. 

The symbol $s_{k}$ is multiplied by a spreading code $\mathbf{c}_{k}^{J_{k}}$ of length $L$ in the time domain. This is indexed by $J_{k} \in \mathcal{J}_{k} = \left\lbrace 1 \cdots N_{c} \right\rbrace$ and indicates the selection of a code in a pool that contains $N_{c}$ of them. The signal is then transmitted via the subcarrier indexed by $I_{k} \in \mathcal{I}_{k} = \left\lbrace 1 \cdots N \right\rbrace $ out of $N$ available subcarrier indices. The spreading code $\mathbf{c}_{k}^{J_{k}} \in \mathbb{C}^{L}$ is selected from a predefined codebook in the block $k$ denoted as $\mathcal{C}_{k} = \left\lbrace \mathbf{c}_{k}^{1}, \cdots  \mathbf{c}_{k}^{N_{c}} \right\rbrace$.

Since part of the transmitted bits are conveyed by the code and the frequency indices, and to have a fair comparison with other conventional systems, we define an equivalent system bit energy $E_{bs}$ which represents the effective energy consumed per transmitted bit. This equivalent system bit energy is related to the physically modulated bit energy by the relationship:
\begin{equation}
E_{bs} = \frac{p}{p_{1}} E_{b}.
\label{eq:1.1.1}
\end{equation}

Thus, the transmitted symbol has an energy per coded bit $E_s = \mathbb{E}[s_{k}s_{k}^{*}] = p_{1}E_{bs}$. At the subcarrier indexed by $I_{k}$, the spread signal $\mathbf{X}_{k}^{(I_{k})}$ is given by:
\begin{equation}
\mathbf{X}_{k}^{(I_{k})} =\left[X_{k,0}^{(I_{k})},\cdots,X_{k,L-1}^{(I_{k})}\right]^{T}  = \left[s_{k}c_{k,0}^{J_{k}},\cdots, s_{k}c_{k,L-1}^{J_{k}}\right]^{T}
\label{eq:1.1.2}  
\end{equation}
while the inactivate subcarriers $\forall i \neq I_{k}$, $\mathbf{X}_{k}^{(i)}$ are zero vectors of $L$ elements. Due to their excellent correlation properties, orthogonal spreading codes such as Walsh-Hadamard or Zadoff-Chu may be used. In this work, we employ Walsh-Hadamard codes. Under such a condition, we have 
\begin{equation}
\mathbf{c}_{k}^{J_{k}} \cdot \left( \mathbf{c}_{k}^{J_{k}^{'}}\right)^{*} = \sum\limits_{l = 0}^{L-1} c_{k,l}^{J_{k}}\left( c_{k,l}^{J_{k}^{'}}\right)^{*}  =
 \left\{
                \begin{array}{ll}
                   1 & \mbox{if}~J_{k} = J_{k}^{'} \\
                  0 & \mbox{otherwise.}\\                 
                \end{array}
              \right.
\label{eq:1.1.2bis}               
\end{equation}
where $(\cdot)^{*}$ is the complex conjugate. 
After obtaining and concatenating $\mathbf{X}_{k}$ for all $k$, we have a vector $\mathbf{X}$
\begin{equation}
\mathbf{X} = \left[\underbrace{X_{1,l}^{(1),}\cdots, X_{1,l}^{(N)}}_{\mathbf{X}_{1}}, \cdots,\underbrace{X_{K,l}^{(1)}, \cdots, X_{K,l}^{(N)}}_{\mathbf{X}_{K}}  \right]^{T},
\label{eq:1.1.3}
\end{equation}
where $X_{k,l}^{(i)}$ for all $k$ is given by
\begin{equation}
X_{k,l}^{(i)} =  \left\{
                \begin{array}{ll}
                   s_{k}c_{k,l}^{J_{k}} & \mbox{if}~i = I_{k} \\
                  0 & \mbox{otherwise.}\\                 
                \end{array}
              \right.
\end{equation}

In addition, the orthogonality between the subcarriers holds is assumed in order to combat the intersymbol interference. This is given by $\Delta f = 1/T_{N} \geq 1/T_{c}$ where $T_{N}$ is the duration of a CFIM symbol, and $T_{c}$ is the chip interval of the spreading code. Note that direct spread spectrum is applied to each active subcarrier. Thus, the spreading operation is performed over time. This process is similar to multicarrier direct-sequence code division multiple access (MC-DS-CDMA) systems.

Afterwards, the remaining procedures are the same as those classical OFDM, where $N_T$ is the length after performing the fast Fourier transform (FFT). So, the complete CFIM transmitter that incorporates all blocks would be stated as:
\begin{equation}
x(t) = \frac{1}{\sqrt{T_{N}}} \sum\limits_{k = 1}^{K}\sum\limits_{i = 1}^{N}\sum\limits_{l = 0}^{L-1} X_{k,l}^{(i)}p(t-lT_{c})e^{j2\pi f_{i,k}t},  
\label{eq:2.1.0.bisss}
\end{equation}
where $p(t)$ is assumed to be a rectangular signaling pulse shifted in time given by
\begin{equation}
p(t) = \left\{
                \begin{array}{ll}
                  1 & \mbox{if}~0 \leq t \leq T_{N}\\
                  0 & \mbox{otherwise.}\\                 
                \end{array}
              \right.
\end{equation}
The subcarrier $f_{i,k}$ is given by
\begin{equation}
f_{i,k} = \frac{N(k-1)}{T_{N}}+ \frac{i}{T_{N}}.
\end{equation}
Note that the active subcarrier index is given by $f_{I_k,k}$, with $i = I_{k}$. 

\begin{figure}[!t]
\centering
\includegraphics[width=2.0in,height=2.0in]{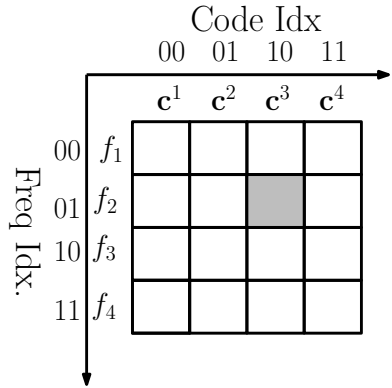}
\caption{An illustration of the CFIM system with $4$ subcarriers and $4$ codes where the transmitter has indexed the message $10$ and $01$. Then encodes the rest of the message via the spreading code $\mathbf{c}_{3}$, transmits the encoded message via the subcarrier $f_{2}$ only.
}
\label{fig:1.1.3.Scheme}
\end{figure}

As a matter of fact, the insertion and removal of cyclic guard prefix is not expressed in our mathematical equations for sake of simplicity. Fig. \ref{fig:1.1.3.Scheme} illustrates the case where $\mathbf{b}_{k}^{p_{2}} = [1,0]$ and $\mathbf{b}_{k}^{p_{3}} = [0,1]$ which leads to the selection of the spreading code $\mathbf{c}_{3}$, and the selection of the second subcarrier, \textit{i.e.} $I_{k} = 2$ out of $4$ available subcarriers in the block.

\subsection{The Receiver}
\label{sect:II.b}
\begin{figure*}[!t]
\centering
\includegraphics[width=7.0in,height=3.0in]{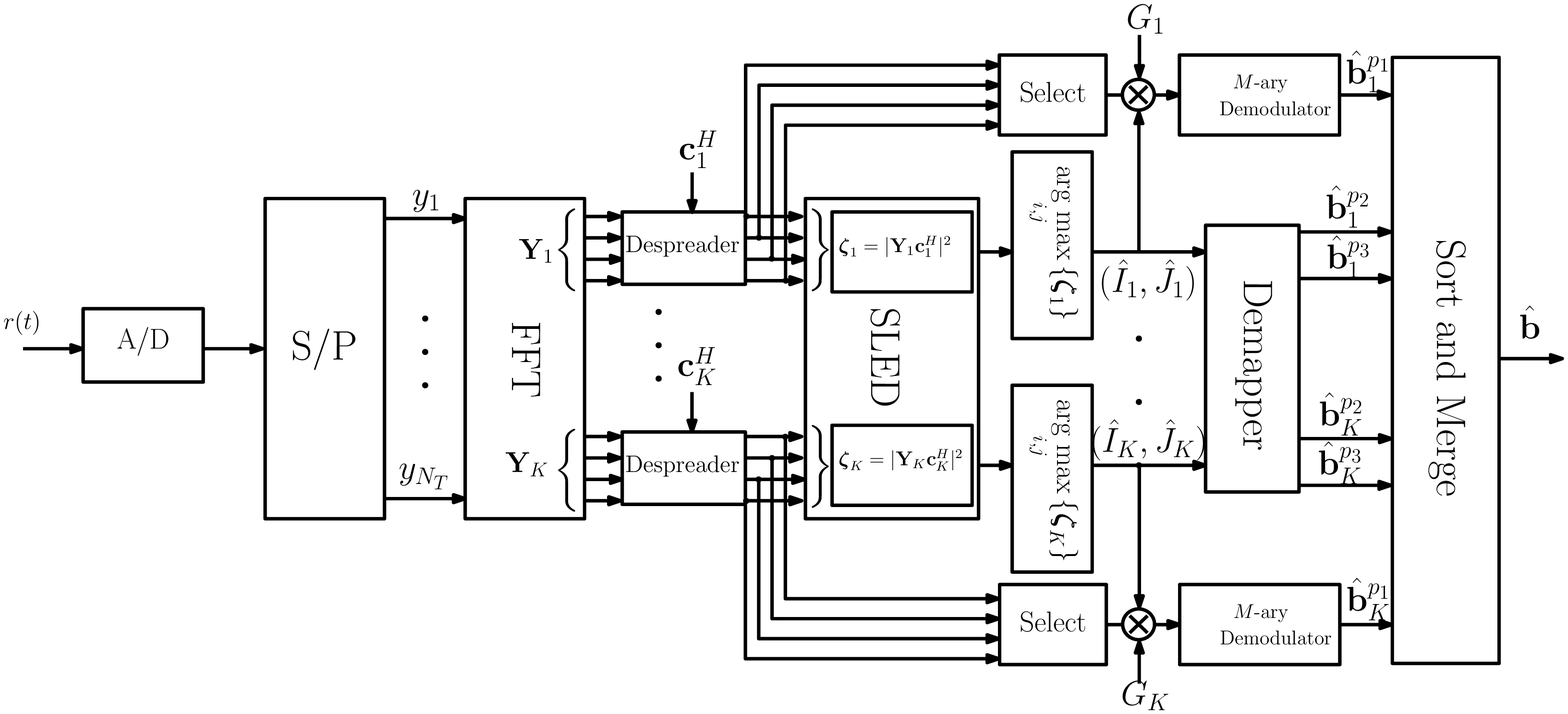}
\caption{CFIM architecture of the receiver}
\label{fig:1.2.CFIM_Rx}
\end{figure*}

A typical architecture of the CFIM receiver with $K \times N$ subcarriers is depicted in Fig. \ref{fig:1.2.CFIM_Rx}. In this paper, we assume that the chip interval $T_{c}$ is larger than the delay spread of the multipath fading channel, so that a flat Rayleigh fading channel is assumed. Moreover, an additive white Gaussian noise (AWGN) is also considered, In this condition, the spread received signal denoted by $Y^{(i)}_{k,l}$ at the $i^{\rm th}$ subcarrier for all $l = 0 \cdots L-1 $ in the $k^{\rm th}$ block is given by
\begin{equation}
Y^{(i)}_{k,l} =  \left\{
                \begin{array}{ll}
                   s_{k} c_{k,l}^{J_{k}}h_{k}^{(i)}+Z_{k,l}^{(i)} & \mbox{if}~i = I_{k} \\
                  Z_{k,l}^{(i)'} & \mbox{otherwise,}\\                 
                \end{array}
              \right.
\label{eq:1.2.1}                            
\end{equation}
where $h_{k}^{(i)} \sim \mathbb{C}\mathcal{N}(0,1)$ is the FFT output produced by the Rayleigh fading channel effect, $ Z_{k,l}^{(i)}$ and $ Z_{k,l}^{(i)'}$ are independent complex AWGN, with zero mean and variance $N_{0}/2$. 

In order to estimate the transmitted message in the $k^{\rm th}$ block with low-complexity decoding, we may first consider a matrix of $N\times L$ elements denoted by $\mathbf{Y}_{k}$ such that
\begin{equation}
\mathbf{Y}_{k} = \left( \begin{array}{ccc}
Y^{(1)}_{k,0}  & \cdots & Y^{(1)}_{k,L-1} \\
\vdots &  & \vdots \\
Y^{(N)}_{k,0} & \cdots & Y^{(N)}_{k,L-1} \end{array} \right)
\label{eq:1.2.1.b}
\end{equation}
Therefore, the receiver uses the a matrix $\mathbf{c}_{k} = \left[\mathbf{c}_{k}^{1},\cdots,\mathbf{c}_{k}^{N_{c}} \right]^{T} $ of $N_{c}\times L$ elements that contains those spreading codes in the codebook $\mathcal{C}_{k}$ to despread the matrix $\mathbf{Y}_{k}$. So, the output of the despreader would simply be
\begin{equation}
\hat{\mathbf{X}}_{k} = \mathbf{Y}_{k}\mathbf{c}_{k}^{H},
\label{eq:1.2.1.c}
\end{equation}
where $(\cdot)^{H}$ is the Hermitian transpose of the matrix. Then, a SLED is used to estimate both the active subcarrier and the selected code indices  in the $k^{\rm th}$ block. Since the spreading codes as well as the subcarriers are mutually orthogonal, the output variables in the matrix $\hat{\mathbf{X}}_{k} $ are fed to SLED forming an $N\times N_{c}$ matrix $\mathbf{\zeta}_{k}$ of decision variables such that for $i = 1,\cdots, N$ and $j = 1,\cdots, N_{c} $ we have
\begin{equation}
\zeta_{k}^{(i,j)} = \big\vert \hat{X}^{(i,j)}_{k} \big\vert^{2} = \left\{ 
                \begin{array}{ll}
                    \big\vert s_{k}h_{k}^{(i)}+\eta_{k}^{(i,j)} \big\vert^{2}& \mathcal{H}_{1}\\
                   \big\vert\eta_{k}^{(i,j)'} \big\vert^{2} & \mathcal{H}_{0},\\                
                \end{array}
              \right.
\label{eq:1.2.2}                            
\end{equation}
where $ \mathcal{H}_{1}$ and $ \mathcal{H}_{0}$ are two hypotheses with which $ \mathcal{H}_{1}$ indicates the presence of the symbol $s_{k}$, and $\mathcal{H}_{0}$ is the alternative that there is no signal. The additive noise component $\eta_{k}^{(i,j)}$ is given by
\begin{equation}
\eta_{k}^{(i,j)} = \sum_{l = 0}^{L-1} Z_{k,l}^{(i)}\left( c^{j}_{k,l} \right)^{*},
 \end{equation} 
and $\eta_{k}^{(i,j)'}$ is written in the same manner as the latter with $Z_{k,l}^{(i)'}$. Since the AWGN components $Z_{k,l}^{(i)}$ and $Z_{k,l}^{(i)'}$ are mutually independent, $\eta_{k}^{(i,j)}$ and $\eta_{k}^{(i,j)'}$ are also independent. 

In order to estimate the indices actually involved in transmission i.e $(I_{k},J_{k})$, the SLED chooses the arguments of the maximum value of the matrix $\mathbf{\zeta}_{k}$ such that 
\begin{equation}
(\hat{I}_{k},\hat{J}_{k}) =  \underset{i\in \mathcal{I}_{k},j\in \mathcal{J}_{k}}{\mbox{argmax}}\left\lbrace \zeta_{k}^{(i,j)}  \right\rbrace. 
\end{equation}

Therefore, the decision variable might land in $\mathcal{H}_{1}$ if $(\hat{I}_{k},\hat{J}_{k}) = (I_{k},J_{k})$. By estimating the active subcarrier and the selected code indices $(\hat{I}_{k},\hat{J}_{k})$, the receiver can extract the $p_{3} + p_{2}$ \emph{mapped} bits. The receiver then demodulates the corresponding branch output using a conventional $M-$ary demodulator to extract the remaining $p_{1}$ \emph{modulated} bits. 
In this paper, we assume that the channel coefficients are perfectly estimated by the receiver. Thus, an equalization is performed at the receiver by dividing the received symbol $\hat{X}_{k}^{(i,j)}$ by $G_{k} = 1/h_{k}^{(i)}$. 

By virtue of SLED, CFIM exhibits a low-complexity structure in comparison to other IM schemes. When compared to OFDM-IM for instance, our method does not require sharing an indexing look up table between communicating parties.

\section{ Performance Analysis}
\label{sect:III}
In this section, we investigate the performance of the CFIM system in terms of the probability of error and review the system gains obtained by concurrently mapping information bits into subcarrier and spreading code indices.
 
\subsection{Probability of Bit Error of the System}
In the CFIM scheme, the transmitted data in every block can be divided into three parts; two blocks to represent the \emph{mapped} bits and a single block to identify the \emph{modulate} bits. The former two blocks determine the combination of subcarrier and spreading code selected, while the latter block contains the remaining data.

In such structure, the probability of bit error of the CFIM system consists of the probability of bit error of the mapped bits $P_{\rm map}$ and the probability of bit error of the modulated bits $P_{\rm mod}$. Subsequently, the probability of bit error of the system could be described as
\begin{equation}
P_{\rm{CFIM}} = \frac{p_{1}}{p}P_{\rm mod}+\frac{p_{2}+p_{3}}{p}P_{\rm{map}}.
\label{eq:3.1.1}
\end{equation}
The probability of errors $P_{\rm mod}$ and $P_{\rm{map}}$ are respectively weighted by the number of modulated bits and the number of mapped bits divided by the total number of bits involved in transmission.

Moreover, these probabilities of error are linked to the probability of erroneous detection of the code-frequency index $P_{\rm ed}$ pair. Indeed, $P_{\rm mod}$ depends on the correct estimation of the selected indices and the probability of bit error of the $M-$ary modulation $P_{b}$ when the indices are correctly estimated. Thereupon, errors happen in two different manners. The first case is when the selected indices are correctly estimated but an error in the demodulation process takes place. The second case is when an error befalls in the assessment of the indices chosen at transmission, and the modulated bits are therefore estimated using inaccurate code-frequency indices. In this case, the receiver has no choice but to guess the modulated bits. In fact, the probability of bit error will be simply equal to $1/2$. Consequently, the BER probability of the modulated bits would be expressed as:
\begin{equation}
P_{\rm mod} = P_{b}(1-P_{\rm ed})+\frac{1}{2}P_{\rm ed}. 
\label{eq:3.1.2}
\end{equation}

In determining the $P_{\rm map}$, the erroneous detection of the code-frequency index pair can cause a wrong estimation of the combination of mapped modulated bits. Each wrong combination can have a different number of incorrect bits compared to the correct combination, i.e the actually transmitted one. 
Since, we have assumed that the frequency and the spreading codes are mutually orthogonal, the detection of the indices is simply equivalent to a noncoherent $N\times N_{c}$-ary orthogonal system upon which operates the SLED. Therefore, the probability of index in error $P_{\rm ed}$ is converted into the corresponding BER probability of mapped bits as
\begin{equation}
P_{\rm map} = \frac{2^{(p_{2}+p_{3}-1)}}{2^{p_{2}+p_{3}}-1}P_{\rm ed}.
\label{eq:3.1.3}
\end{equation} 

\subsection{Probability of Erroneous Detection of the Code-Frequency Index}
Since a SLED is used for estimating the pair of indices, we shall determine the probability of erroneously detecting the pair of the code-frequency indices $P_{\rm ed}$. To do so, we assume an equiprobable selection of the active subcarrier and the spreading code. Moreover, for the sake of clarity, we focus on a single block, \textit{i.e} the $k^{\rm th}$ block such that $\zeta_{k}^{(i,j)}$ of (\ref{eq:1.2.2}) simplifies to $\zeta^{(i,j)}~\forall i \in \mathcal{I}$ and $\forall j \in \mathcal{J}$. Moreover, Since the frequency as well as the spreading codes are mutually orthogonal, we can write the matrix $\boldsymbol{\zeta}$ as a vector of $NN_{c}$ elements, which is denoted as $\mbox{\textit{vec}}(\boldsymbol{\zeta}) = \left[\zeta^{(1,1)},\cdots, \zeta^{(N,N_{c})}\right] = \left[\zeta^{(1)},\cdots, \zeta^{(NN_{c})}\right]$. Denoting the selected index at the transmitter as $\nu$, the probability of index error $P_{\rm ed}$ conditioned on the selected index $\mu \in \left\lbrace 1,\cdots, NN_{c} \right\rbrace $ would be
\begin{equation}
\begin{split}
P_{\rm ed} &= P\left\lbrace \zeta^{(\nu)} < \max\left( \zeta^{(\mu)} \right)\mid \nu \right\rbrace,\\
& ~~~~\mbox{for}~1 \leq \underset{\mu \neq \nu}{\mu} \leq NN_{c}.  
\end{split}
\label{eq:3.2.1}
\end{equation}
As deduced from in equation (\ref{eq:3.2.1}), an error in the estimation of the selected index will occur if the decision variable $\max\left( \zeta^{(\mu)} \right)$ is larger than $\zeta^{(\nu)}$. In this condition, it is easy that $P_{\rm ed}$ is equivalent to the probability of detection of a noncoherent $N\times N_{c}$-ary orthogonal system over the Rayleigh channel. This is given by \cite{Proakis2001}:
\begin{equation}
P_{\rm ed} = \sum^{NN_{c}-1}_{\mu = 0} \frac{(-1)^{\mu}}{1+\mu + \mu \bar{\gamma_{c}}} \binom{NN_{c}-1}{\mu}, 
\end{equation} 
where $\bar{\gamma_{c}} = \mathbb{E} \left[ \vert h \vert ^{2}\right] \frac{E_{s}}{N_{0}}$ is the average signal-to-noise ratio (SNR) per symbol.

\subsection{Probability of Bit Error of Modulated Bits}
When the subcarrier and spreading code indices are correctly estimated, then we must consider the presence of the channel coefficient $h$ when formulating the average bit error probability $P_b$ of the modulated bit, which becomes conditional on the received power and may be drafted as
\begin{equation}
   P_{b} = \int_{0}^{+\infty} P_{b\mid \gamma_{b}} P(\gamma_{b}) d\gamma_{b},
   \label{eq:3.3.1}
   \end{equation}   
where $\gamma_{b} = \vert h \vert^{2}\frac{E_{bs}}{N_{0}}$ is the instantaneous SNR per bit. $P(\gamma_{b})$ is the conditional probability distribution function of $\gamma_{b}$ given the correct index estimation. $P_{b\mid \gamma_{b}} $ is the conditional probability of bit error in AWGN channels \cite{Simon2005}. Specifically, for M-PSK modulation using Gray code, we may use the close expression in \cite{Lee1986} to derive $P_{b\mid \gamma_{b}}$:
 \begin{equation}
\begin{split}
   P_{b\mid \gamma_{b}}   \approx \frac{1}{p_{1}} \sum_{u = 1}^{M/2} w_{m}^{'} P_{m},
\end{split}
 \label{eq:3.3.2}
\end{equation}
where $w_{m}^{'} = w_{m}+w_{M-m}$, $w_{M/2}^{'} = w_{M/2}$, $w_{m}$ is the Hamming weight of bits assigned to the symbol $m$, and $P_{m}$ is given in equation (\ref{eq:3.3.3}) as stated in \cite{Lee1986}.

\begin{figure*}[!t]
\begin{equation}   
P_{m} = \frac{1}{2\pi} \left\lbrace \int\limits_{0}^{\pi(1-(2m-1)/M)} \hspace{-12pt} \exp\left( - p_{1}\gamma_{b}\frac{\sin^{2}\left[(2m-1)\pi/M \right] }{\sin^{2}\theta}\right)d\theta- \hspace{-10pt} \int\limits_{0}^{\pi(1-(2m+1)/M)} \hspace{-12pt} \exp\left( - p_{1}\gamma_{b} \frac{\sin^{2}\left[(2m+1)\pi/M \right] }{\sin^{2}\theta}\right)d\theta\right\rbrace
\label{eq:3.3.3} 
  \end{equation}  
  \vspace*{-2pt}
\hrulefill
\vspace*{2pt}
\end{figure*}

In order to compute the probability of bit error $P_{b}$, the probability distribution of $\gamma_{b}$ needs to be derived. It should be noted that the random variable $\gamma_{b}$ is obtained knowing that the index estimation is correct. As a result, if the index estimation is always correct, \textit{i.e.} $P_{\rm ed} = 0$, then $\gamma_{b}$ is chi-square distributed with two degrees of freedom. This is due to the fact that $h \sim \mathbb{C}\mathcal{N}(0,1)$. In this case, $P_{b}$ in equation (\ref{eq:3.3.1}) is equivalent to the probability of bit error of the conventional MPSK in Rayleigh fading channels. However, in the low-SNR regime, SLED selects the index that has the greatest SNR. Consequently, $\gamma_{b}$ is no longer chi-square distributed with two degrees of freedom. In such a condition, the conditional probability distribution function $P(\gamma_{b})$ is not available. Nevertheless, an empirical distribution may be obtained. 

By computing $P_{b}$ using equations (\ref{eq:3.3.2}) and \eqref{eq:3.3.3}, the BER probability of the modulated bits $P_{\rm mod}$ can be obtained via equation  \eqref{eq:3.1.2}. With $P_{\rm map}$ in equation \eqref{eq:3.1.3}, the comprehensive probability of bit error of the proposed $P_{\rm{CFIM}}$ scheme is presented in equation \eqref{eq:3.1.1}. 

 
\section{Spectral Efficiency, Energy Efficiency and Complexity Analyses}
\label{sect:IV}
In this section, we consider the spectral and the energy efficiency as well as the complexity analyses for the proposed CFIM system.
\subsection{Spectral Efficiency}
The CFIM system has a total of $KN$ subcarriers, but uses $K$ subcarriers to transmit $Kp$ bits at every transmission instant. In fact one out of $N$ subcarriers is activated in each block, so the spectral efficiency is simply indicated as
\begin{equation}
\xi_{\rm CFIM} = \frac{p}{N} = \frac{p_{1}+p_{2}+p_{3}}{2^{p_{2}}}.
\label{eq:SPE_EFF}
\end{equation}

It can be seen that increasing $p_2$ e.g the number of subcarriers would reduce the spectral efficiency, while increasing the modulation order and the number of spreading codes both enhance the spectral efficiency. However, these would impact the overall performance of the system.

\subsection{Energy Efficiency}
In CFIM systems, only $Kp_{1}$ bits from the total $Kp$ are directly modulated using $M-$ary modulation, whereas $K(p_{2}+p_{3})$ bits are conveyed in the selection of codes and subcarriers. Considering that each modulated bit requires an energy of $E_b$ to be transmitted, then mapping to index subcarriers and spreading codes should reduce the total required transmission energy. Consequently, the percentage of energy saving per block in the proposed system is given by
\begin{equation}
\begin{split}
E_{\rm saving} &= \left(1-\frac{p_{1}}{p} \right) E_{b} \%\\
& =  \left(1-\frac{1}{1+\frac{p_{2}+p_{3}}{p_{1}}} \right) E_{b} \%.
\end{split}
\label{eq:NRG_EFF}
\end{equation}
Under this condition, the energy saving depends on the ratio of the number of subcarriers and spreading codes involved in indexing to the modulation order of the system. Clearly, an augmentation in the number of subcarriers or spreading codes results in more energy saving in the CFIM system. However, increasing the number of subcarriers reduces the spectral efficiency. Hence, indexing via spreading codes is associated with less cost in terms of energy and spectral efficiency.

A drawback of multicarrier systems is their high peak to average power ratio (PAPR). This impacts the performance by distorting the signal induced by the nonlinearity of high power amplifier (HPA) \cite{Jiang2013,Jiang2008,Armstrong2002}. This could in turn deteriorate the spectral and energy efficiency of the system. Reducing PAPR leads to a significant power saving, which improves the energy efficiency performance. In fact, a high PAPR value appears when a number of subcarriers in a given OFDM system are out of phase with each other. Thus, a high PAPR value occurs when a large number of subcarriers is activated. Since the proposed scheme activates a single frequency in each block, PAPR is reduced. The PAPR is defined as
\begin{equation}
\begin{split}
\rm{PAPR} &= \frac{ \underset{ 0 \leq t \leq T_N}{\max } \vert x(t) \vert^{2}}{1/T_N \int_{0}^{T_N} \vert x(t) \vert^{2} dt}.\\
\end{split}
\label{eq:PAPR}
\end{equation}
Using equation \eqref{eq:2.1.0.bisss}, the maximum expected PAPR can be determined. In conventional OFDM, assuming that all symbols are equal, the peak value is given by
\begin{equation}
 \max\left[ x(t)x^{*}(t) \right] =  \frac{E_s}{T_N} K^{2} N^{2}L^{2}, 
 \end{equation} 
while the mean square value of the signal is
\begin{equation}
 \mathbb{E}\left[ x(t)x^{*}(t) \right] =  \frac{E_s}{T_N} K NL. 
 \end{equation} 
In this condition the maximum expected PAPR of the OFDM is $K N L$. In CFIM, since a single subcarrier per block is activated, it is easy to determine the maximum expected PAPR as $K L$. Therefore, the CFIM can reduce the PAPR by $1/N$ compared to conventional OFDM. Since $N = 2^{p_2}$, the PAPR will significantly reduce as the number of bits $p_{2}$ increases.

\subsection{System Complexity}
The complexity of the CFIM system can be evaluated by the number of operations required to accomplish transmission. Since the CFIM uses IFFT and FFT operations with a length of $N_{T}$, the computational complexity will be $\mathcal{O}_{\text{FFT/IFFT}} \sim \mathcal{O}(2N_{T}\log_{2}(N_{T}))$. For the sake of clarity, this will be omitted in this paper. 

The number of spreading codes and not its length is used for indexing. Considering that each spreading code contains $L$ elements, and the transmitter selects one  out of $N_{c}$ codes in the $k^{\rm th}$ block, the resulting number of operations is $L$ on the transmitter side.

The receiver forms the matrix  $\mathbf{Y}_{k} \in \mathbb{C}^{N \times L} $ as described in equation (10) after the FFT operation.
In order to estimate the transmitted information, the receiver despreads the received signal using a matrix $\mathbf{c}_{k}$ of $N_{c}\times L$ elements as in equation \eqref{eq:1.2.1.c}. As such, the computational complexity of
this operation becomes $\mathcal{O}_{\text{spread/despread}} \sim \mathcal{O}(2NLN_{c}-NN_{c}+L)$. Since there are $K$ blocks, the overall complexity 
that incorporates the transmission and reception turns into
$K\mathcal{O}_{\text{spread/despread}}$.

The SLED multiplies the vector $\hat{\mathbf{X}}_{k}$ by its complex conjugate. Therefore, the computational complexity is simply $\mathcal{O}_{\text{SLED}} \sim \mathcal{O}(N N_{c}K)$. Regarding the modulation and the demodulation, we assume that a modulator converts a stream of $p_{1}$ bits into an $M-$ary symbol by computing an inner product between the bit stream and the vector $\left[2^{p_{1}-1}, \cdots, 1 \right]$. The demodulator will convert the symbol into a bit stream by computing $p_{1}$ Euclidean divisions. As a result, the computational complexity of a modulator and a demodulator will be $\mathcal{O}_{\text{Mod/Demod}} \sim \mathcal{O}(3p_{1}-1)$ per active subcarrier. In total, the complexity of the CFIM system would be expressed as
\begin{equation}
\mathcal{O}_{\text{CFIM}} = \mathcal{O}_{\text{SLED}}+K(\mathcal{O}_{\text{spread/despread}}+\mathcal{O}_{\text{Mod/Demod}}).
\label{eq:O_CFIM}
\end{equation}

In order to have a fair comparison between the proposed system with other multi-carrier systems in the state of the art, we assume that direct spread spectrum communications are used in the following systems. In the conventional DS-OFDM scheme, all subcarriers are activated during the transmission, and a spreading code is used in each subcarrier. Since there is one spreading code per subcarrier, the complexity of the spreading and despreading operations is given by $\mathcal{O}_{\text{spread/despread}} \sim \mathcal{O}(2L-1+L)$ per subcarrier. Assuming that there is $K$ different symbols that are distributed over $KN$ subcarriers, the computational complexity will be
\begin{equation}
\mathcal{O}_{\text{DS-OFDM}} = KN(\mathcal{O}_{\text{spread/despread}}+\mathcal{O}_{\text{Mod/Demod}}).
\label{eq:O_MC_DS_CDMA}
\end{equation} 
Aiming at comparing the CFIM system to systems which share the same modulation concept but with a different receiver design, we compare the complexity of CFIM to that of DS-OFDM-IM. In the latter, only $g$ out of $N$ subcarriers are activated. However, the complexity of the maximum likelihood detector is $\mathcal{O}_{\text{MLD}} \sim \mathcal{O}(2^{p_{2}} M^{g})$ per active subcarrier where $g = p_{2}/\log_{2}(M)$, \cite{Basar2013}. 
Since $g$ spreading codes are used per block, the computational complexity is $gL$ per active subcarrier. 
At the receiver, despreading of the received signal is required. Since there are $g$ activated subcarriers out of $N$, there exists $N$ choose $g$ times $g!$ possible combination. Under this condition the computational complexity that incorporates spreading and depsreading operations would be $\mathcal{O}_{\text{spread/despread}}\sim {N \choose g} g! (2L-1)+gL$. Assuming that there are $K$ different blocks the total computational complexity of the system becomes    
\begin{equation}
\begin{split}
\mathcal{O}_{\text{OFDM-IM-DS-CDMA}} = K(g\mathcal{O}_{\text{spread/despread}}+\mathcal{O}_{\text{Mod/Demod}})+\mathcal{O}_{\text{MLD}}.
\end{split}
\label{eq:O_OFDMIM_DS_CDMA}
\end{equation} 

\section{Extension to Synchronous Multiuser Communications}
\label{sect:V}
In this section, we extend the proposed system to synchronous multiuser communications.
   
\subsection{Index Modulation for Multiuser Communications using CFIM}
\begin{figure}[!h]
\centering
\includegraphics[width=2.7in,height=2.0in]{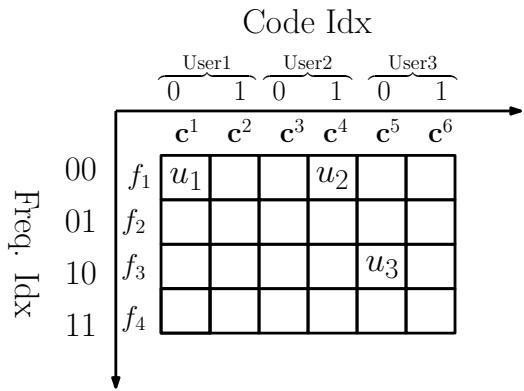}
\caption{The CFIM system with $4$ subcarriers and $2$ codes per user. In the illustration, $00$ has been indexed for User $1$ and User 2, and thus $f_{1}$ has been selected, while $f_{3}$ has been selected for User $3$. Regarding the spreading code selection, $\mathbf{c}_{1}$, $\mathbf{c}_{4}$ and $\mathbf{c}_{5}$ have been selected respectively for User $1$, $2$ and $3$.}
\label{fig:4.0.1.Scheme}
\end{figure}

Assuming that a codebook $\mathcal{C}_{k}$ in the $k^{\rm th}$ block can be split into a group of $U$ codebooks of $N_{c}$ orthogonal spreading codes each. In this condition, CFIM can be extend to multiuser communications where it operates in the same frequency band with no multiuser interference (MUI). Fig. \ref{fig:4.0.1.Scheme} illustrates the case where three users have two spreading codes each and four subcarriers are available for transmission. 

\subsection{Downlink Transmission}
\begin{figure}[!t]
\centering
\includegraphics[width=3.2in,height=1.2in]{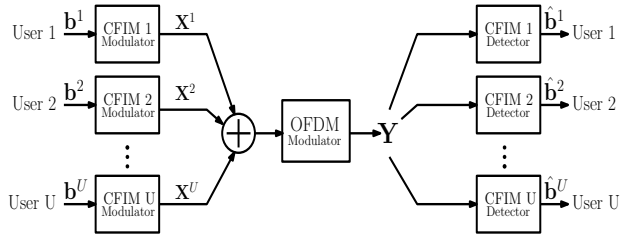}
\caption{Multiuser CFIM in downlink scenario}
\label{fig:4.1.1.DL}
\end{figure}

In downlink transmission, a base station prepares one message for each user using $U$ CFIM modulators and transmits the overall signal where they are expected to be received by $U$ independent receivers. In synchronous transmission, MUI can be avoided by using orthogonal spreading codes. Fig. \ref{fig:4.1.1.DL} illustrates the multiuser CFIM in downlink scenario. 

Focusing on the $k^{\rm th}$ block, the spread received signal at the $i^{\rm th}$ subcarrier for all $l = 0\cdots L$ is given by
\begin{equation}
Y^{(i)}_{k,l} =  \left\{
                \begin{array}{ll}
                 h_{k}\sum\limits_{u=1}^{N_{u,i}} s_{k}^{u}c_{k,l}^{J_{k,u}}+Z_{k,l}^{(i)} & \mbox{if}~i = I_{k,u} \\
                  Z_{k,l}^{(i)'} & \mbox{otherwise,}\\                 
                \end{array}
              \right.   
 \end{equation} 
where $h_{k}$ is the fading channel coefficient, $N_{u,i}$ is the total number of users whose combined signal is transmitted by the $i^{\rm th}$ subcarrier. More, $s_{k}^{u}$,  $c_{k,l}^{J_{k,u}}$ for $l = 0 \cdots L-1$ are respectively the modulated symbol and the spreading code selected by the $u^{\rm th}$ user. Note that $(I_{k,u},J_{k,u})$ is the code-frequency index pair selected by the $u^{\rm th}$ user. $Z_{k,l}^{(i)}$ and $Z_{k,l}^{(i)'}$ are the additive noise component in the $i^{\rm th}$ subcarrier. In synchronous downlink scenario, since orthogonal spreading codes are used, the detection of the transmitted signal for the $u^{\rm th}$ receiver is the same as a single CFIM receiver as described in Section {\ref{sect:II.b}}. 

\subsection{Uplink Transmission}
\begin{figure}[!t]
\centering
\includegraphics[width=3.2in,height=1.8in]{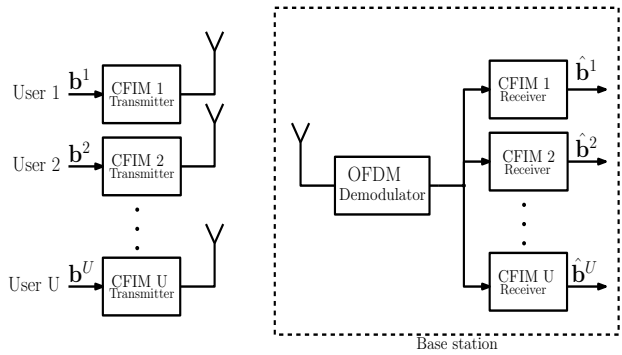}
\caption{Multiuser CFIM in uplink scenario}
\label{fig:4.1.1.UP}
\end{figure}

In uplink transmission, the users transmit $U$ messages to the base station using CFIM transmitters. The base station has $U$ CFIM receivers to decode those messages as depicted in Fig. \ref{fig:4.1.1.UP}. Since we have assumed synchronous transmission, at the $k^{\rm th}$ block, and at the $i^{\rm th}$ subcarrier for all $l = 0\cdots L$ the spread received signal at the base station is given by
\begin{equation}
Y^{(i)}_{k,l} =  \left\{
                \begin{array}{ll}
                 \sum\limits_{u=1}^{N_{u,i}} h_{k,u}s_{k}^{u}c_{k,l}^{J_{k,u}}+Z_{k,l}^{(i)} & \mbox{if}~i = I_{k,u} \\
                  Z_{k,l}^{(i)'} & \mbox{otherwise,}\\                 
                \end{array}
              \right.   
 \end{equation} 
where $h_{k,u}$ is the fading channel coefficient of the uth user. In synchronous transmission, each receiver decodes their message as described in section \ref{sect:II.b}.  

\section{Numerical Results}
\label{sect:VI}
In this section, we study the obtained analytical and simulation results for the proposed CFIM system and show that these results are in good agreement. Then we compare the performance of CFIM to other index-based schemes like SM and OFDM-IM systems. We also study the complexity, spectral and energy efficiency for the proposed system. Finally, CFIM for multiuser communication in synchronous transmission is analyzed. In this paper, we have used a Walsh-Hadamard matrix with various size for spreading operations. Moreover, we have omitted the cyclic prefix for sake of simplicity. 

\subsection{Performance of CFIM}

To have a better sight into the problem, we scrutinize for various scenarios the parameters that affect the performance of the proposed CFIM in this subsection. As the outgoing bits in the system constitute of three main parts that are related to the modulation order $M$, subcarriers $N$ and spreading codes $N_c$, we shall consider the influence of each of these elements on the system performance.

\begin{figure}[!t]
\centering
\includegraphics[width=3.6in,height=2.8in]{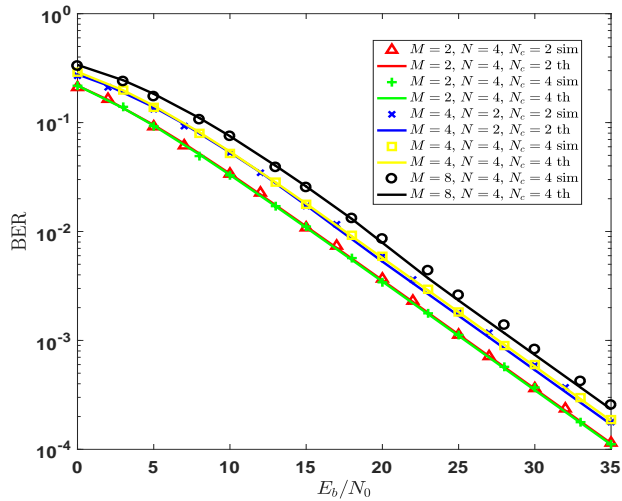}
\caption{Performance of CFIM over Rayleigh fading channels with a spreading factor of $L = 32$. } 
\label{fig:5.1.1}
\end{figure}

We start by plotting the bit error rate (BER) performance of the CFIM scheme for various $M$, $N$ and $N_c$ in Fig. \ref{fig:5.1.1}. This overall performance is extracted from equation \eqref{eq:3.1.1}. 
We witness in this figure that analytical and simulation results for the CFIM system are in correspondence and great harmony, which approves the sureness of our method.
The modulation order $M$ having the most destructive influence on the BER performance is natural, as increasing $M$ reduces the Euclidean distance between the transmitted symbols and narrows down decision zones at the receiver.

Furthermore, Fig. \ref{fig:5.1.1} also shows that our system is design-flexible and adaptable to acquire several forms in terms of $M$, $N$ and $N_c$.  We observe that the CFIM system using $M = 2$, $N = 4$, $N_c = 2$ plotted as red curve and the CFIM using $M = 4$, $N = 2$, $N_c = 2$ plotted as blue curve transmit both $4$ bits per transmission. However, the first one exhibits a better performance, because the modulated part transmits one single bit per symbol while the other transmits two bits per symbol. 

\begin{figure}[!t]
\centering
\includegraphics[width=3.6in,height=2.8in]{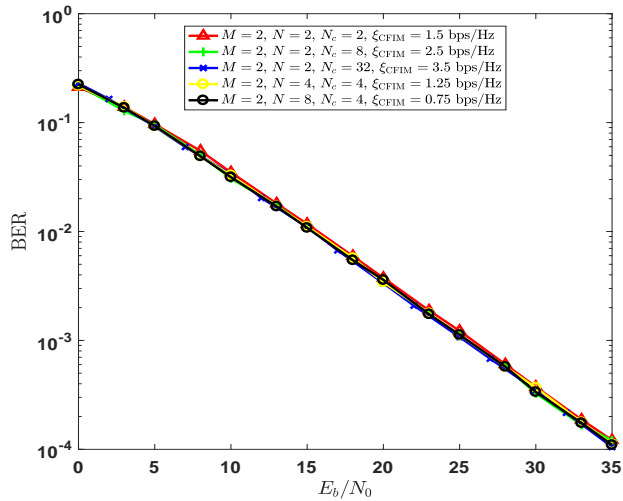}
\caption{Performance of CFIM with a modulation order of $M = 2$, a spreading factor of $L = 32$ and various number of spreading codes and subcarriers.}
\label{fig:5.1.2}
\end{figure}

The BER performance of CFIM for a fixed modulation order $M=2$ and various $N$ and $N_c$ is shown in Fig. \ref{fig:5.1.2}. as depicted in this figure,  the parameters $N$ and $N_c$ have a marginal influence on the BER performance compared to the parameter $M$. In fact, the proposed system exhibits the same property as FIM systems. The BER performance would indeed deteriorate if the contribution made by the energy contained in the mapped bits were not there. This is because higher values of $N$ and $N_c$ would simply challenge the receiver to choose a correct code-frequency index pair from within a larger set and would deteriorate the BER performance. But mapped bits balance the performance by virtue of their energy contribution.

It should be noted that unlike FIM systems, the proposed system can improve the reliability while enhancing SE. Indeed, in FIM schemes only one subcarrier out of $N$ is activated. To get additional mapped bits, this system needs to increase the number of subcarriers available $N$, which requires larger bandwidth expansion, which reduces SE. Fortunately, the proposed CFIM compensates the spectral efficiency loss by indexing via spreading codes.

\subsection{Performance Comparison with FIM, OFDM-IM and SM Systems}
We compare the performance of our proposed work to the performance of two other index modulation based schemes, namely the FIM, the orthogonal frequency division multiplexing with index modulation (OFDM-IM) and the spatial modulation (SM) techniques. The FIM system the mapped bits are indexed by the activation of one out of $N$ subcarriers \cite{Soujeri2017}. The concepts of OFDM-IM and SM are very well explained in \cite{Basar2013} and \cite{Mesleh2006,Mesleh2008,Jeyadeepan2008,Renzo2014}. 

For simplicity, we consider a comparison with these index modulation schemes for the case of dispatching $4$ bits per transmission where the number of subcarriers is limited to $N = 4$, and the number of transmitting antennae for SM is also limited to $N_t = 2$. In this case, since $N = 4$, the CFIM uses a modulation order of $M = 2$, and the number of spreading codes available is $N_c = 2$. In OFDM-IM, we activate $g = 2$ subcarriers out to $N$ and a modulation order of $M = 2$ is used. In SM, since $N_t = 2$, a modulation order of $M = 8$ and a single receiving antenna $N_r = 1$ are considered. In the last two systems, maximum-likelihood is implemented in the receiver to ensure a higher reliability. For the FIM system, a modulation order of $M = 4$ is employed. In addition to these, a comparison with a single carrier system that employs $16$-PSK modulation is considered.   

\begin{figure}[!t]
\centering
\includegraphics[width=3.6in,height=2.8in]{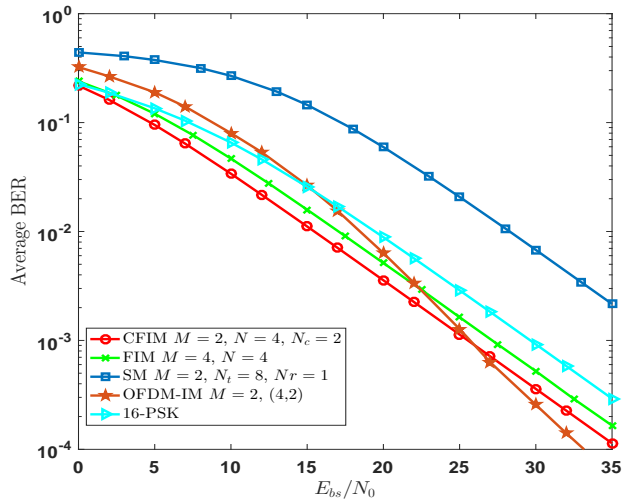}
\caption{Performance of CFIM in comparison to FIM, SM, OFDM-IM, OFDM and $16$-PSK schemes for $4$ bits per transmission}
\label{fig:5.2.1}
\end{figure}

Fig. \ref{fig:5.2.1} shows the performance of the proposed system in comparison to other index modulation based schemes. It can be seen that CFIM outperforms SM, FIM  and single carrier $16$-PSK systems. Although OFDM-IM outperforms other index modulation based systems at $E_b/N_{0} = 25$ dB, CFIM exhibits a better BER performance in the low SNR regime. In IoT applications as well as in wireless sensor networks, communication takes place more often via transmission at the low SNR regime, i.e the SNR would typically range from $0$ to $25$ dB. As a result, CFIM suggests an ideal solution for this type of communication networks, where low transmission power consumption is an absolute requirement.

When the number of subcarriers is limited but we wish to enhance the spectral efficiency, the modulation order needs to be increased in conventional OFDM. In such a condition, the BER performance degrades. The proposed system is more flexible so that it can be designed in different fashions, i.e in a way to optimize the spectral efficiency while having high reliability without any additional bandwidth. 

\begin{figure}[!t]
\centering
\includegraphics[width=3.6in,height=2.8in]{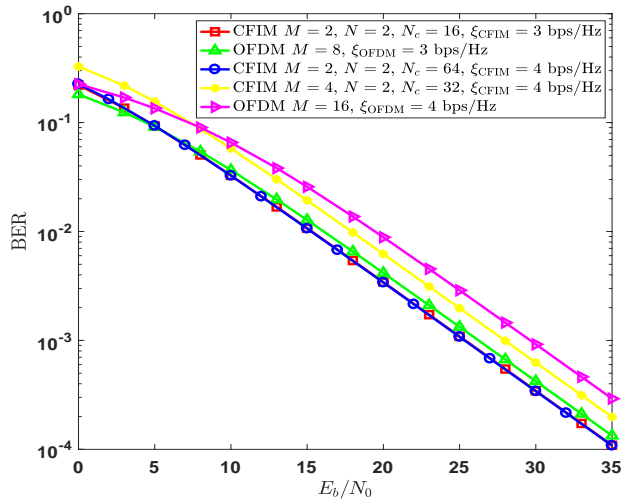}
\caption{Performance comparison of CFIM to conventional OFDM with spectral efficiencies of $3$ bps/Hz and $4$ bps/Hz. Number of blocks is $K = 2$ and number of subcarriers per block is $N = 2$.}
\label{fig:5.2.2}
\end{figure}

In Fig. \ref{fig:5.2.2}, we illustrate this by plotting the performance of CFIM vs. conventional OFDM schemes with spectral efficiencies of $3$ bps/Hz and $4$ bps/Hz. In either cases, with a total number of four subcarriers, it can be seen that CFIM achieves better performances compared to OFDM. In fact, the proposed scheme activates two subcarriers out of four and the number of transmitted bits is maximized with the help of spreading codes selection. This allows for minimizing the modulation order that guarantees higher reliability. Unlike SM and OFDM-IM that use ML detection, such a scheme achieves a noticeable performance despite its simplicity and ease. The proposed system overcomes SM as it does not require neither space nor heavy hardware and it does not totally depend on channel coefficients to perform. It also overcomes OFDM-IM which is far too complex for IoT applications.

\subsection{Complexity, Energy and Spectral Efficiency}
Despite its simplicity, CFIM can achieve better spectral efficiency while maintaining higher reliability compared to other index modulation based systems. Moreover, CFIM as well as FIM schemes activate one subcarrier out of $N$ subcarriers in a block, yielding low-complexity and high energy efficiency communication systems. Therefore, these two criteria are paramount for IoT devices as well as for low-cost wireless sensors. 

\subsubsection{System Complexity}
The computational complexity of these systems is considered for evaluation here. Since, CFIM can be applied to multiuser communications, we apply direct spread spectrum to other communication schemes in order to have a fair comparison. As a result, the complexity analysis of these systems may be discussed by considering DS-FIM, conventional DS-OFDM and DS-OFDM-IM. It should be noted that we have omitted the complexity of the IFFT/FFT part because the number of operations is the same for all the aforementioned systems. 

\begin{figure}[!t]
\centering
\includegraphics[width=3.6in,height=2.8in]{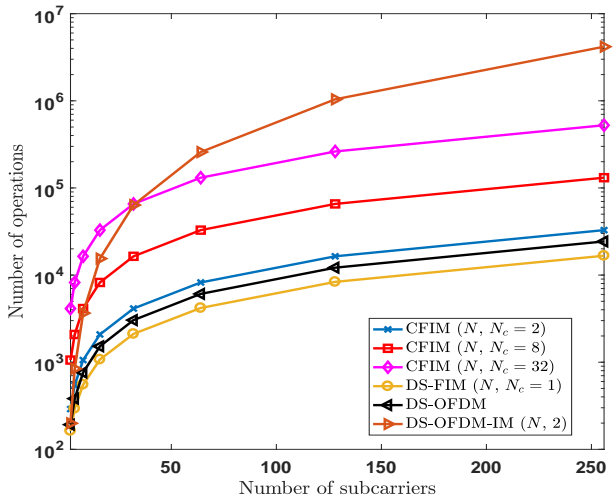}
\caption{Complexity comparison of the proposed CFIM to FIM, OFDM and OFDM-IM($N$,$2$) systems using a modulation order of $M = 2$ and a spreading factor of $L = 32$ in a single block, e.g $K = 1$.}
\label{fig:5.3.1}
\end{figure}

In fact, the complexity of the CFIM depends on the number of spreading codes involved in indexing. This is because on the receiver side, $N_c$ despreading operations are needed on each subcarrier to estimate the mapped bit. Moreover, the SLED has $N\times N_c$ decision variables, which results in more operations when $N_c$ is increased. Therefore, to offset better spectral efficiency and high reliability communication, CFIM may incur more complexity on the receiver side. Fig. \ref{fig:5.3.1} depicts the complexity of the CFIM in comparison with other index modulation based systems according to equations \eqref{eq:O_CFIM}, \eqref{eq:O_MC_DS_CDMA} and \eqref{eq:O_OFDMIM_DS_CDMA}. Note that, the complexity of FIM is similar to equation \eqref{eq:O_CFIM} with $N_c = 1$. 

By fixing the modulation order to $M = 2$ and increasing the number of subcarriers available, FIM exhibits the lowest complexity because only one subcarrier is activated for the transmission, and one single unique spreading code is used for the spreading and the despreading operation. In DS-OFDM, assuming that all subcarriers are activated, more operations is needed to estimate the transmitted information, which results in higher complexity. On the other hand, CFIM systems exhibit higher complexity as the number of spreading codes involved in indexing increases. An augmented system complexity is the price to pay to get a higher number of bits per transmission while maintaining reliability. Nevertheless, the complexity is reasonable for a small number of spreading codes, i.e $N_c = 2$.

Furthermore, it can be seen that the proposed system is clearly the winner compared to OFDM-IM for large $N$ as the latter employs maximum-likelihood detection to estimate the information, with a complexity of $\mathcal{O}_{\text{MLD}} \sim \mathcal{O}(2^{p_{2}} M^{g})$. This induces further complexity when the number of activated subcarriers $g$ is higher.     

\subsubsection{Energy Efficiency}
In order to complete the analysis, we have evaluated the energy efficiency of the proposed CFIM  system in comparison with the aforementioned systems. Since modulated bits are only transmitted via the channel, one can save significant amount of energy by mapping more bits. According to equation \eqref{eq:NRG_EFF}, the system can achieve better energy efficiency by mapping more information bits. However, saving energy comes at the cost of diminished spectral efficiency in FIM system. This is due to the fact that only one single subcarrier is activated out of $N$. 

The proposed CFIM system can save a great amount of energy while maintaining a desired spectral efficiency. This is made possible by indexing via spreading codes. Indeed, this does not require neither additional bandwidth nor energy $E_b$ to transmit extra bits in the system. Table \ref{table:OFDMvsOFDMIMvsFIM} compares our CFIM system with DS-FIM, DS-OFDM and DS-OFDM-IM in terms of complexity, spectral efficiency and energy saving. By using a modulation order of $M = 2$ and $N = 4$ subcarriers in all systems, CFIM is superior in terms of energy saving. Note that increasing the number of spreading codes saves energy, enhances spectral efficiency, but increases complexity. Furthermore, by comparing CFIM($4$,$2$), DS-OFDM and DS-OFDM($4$,$2$) at equal spectral conditions, CFIM exhibits less complexity compared to DS-OFDM($4$,$2$) and saves more energy compared to the two aforementioned systems. 

\begin{table}[!t]
\centering
\caption{Comparison CFIM, DS-FIM, DS-OFDM and DS-OFDM-IM systems using a modulation order of $M = 2$, $N = 4$ subcarriers, and spreading factor $L = 32$}
\label{table:OFDMvsOFDMIMvsFIM}
\renewcommand{\arraystretch}{1.8}
\begin{tabular}{|c|c|c|c|}
\hline
System & Complexity & Spectral Eff.& Energy saving \\
 \hline
CFIM($4,2$)  &  $ 546$ & $1$  bps/Hz & $75\%$ \\
\hline
CFIM($4,8$)   & $2082$ &  $1.5$  bps/Hz&  $83.3\%$\\
\hline
CFIM($4,32$)& $ 8226$ &  $2$  bps/Hz&  $87.5\%$\\
\hline
 DS-FIM & $290$ &  $0.75$  bps/Hz&  $66.7\%$\\
 \hline
 DS-OFDM & $388$ &  $1$  bps/Hz& $0\%$\\
  \hline
  DS-OFDM-IM($4$,$2$)& $841$ &  $1$  bps/Hz& $50\%$\\
   \hline
\end{tabular}
\end{table}

\begin{figure}[!htpb]
\centering
\includegraphics[width=3.6in,height=2.8in]{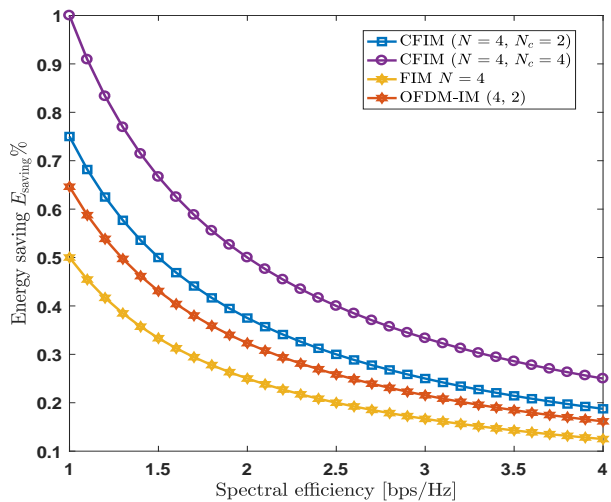}
\caption{Energy efficiency analysis vs. spectral efficiency with $N = 4$ subcarriers}
\label{fig:5.3.2}
\end{figure}

\begin{figure}[!t]
\centering
\includegraphics[width=3.6in,height=2.8in]{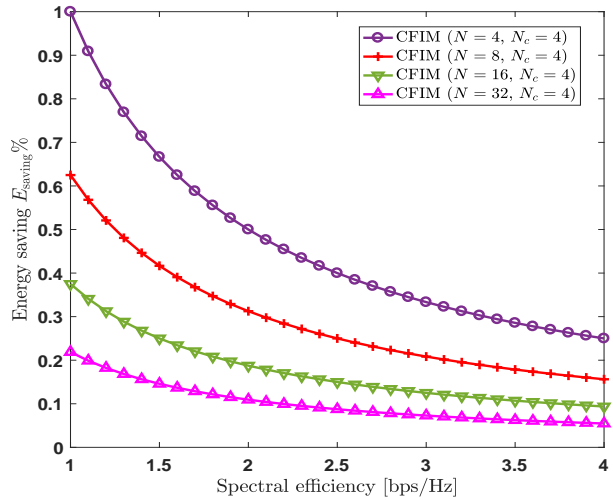}
\caption{Energy efficiency vs spectral efficiency of CFIM with various number of subcarriers and $N_c=4$ spreading codes.}
\label{fig:5.3.3}
\end{figure}

By comparing the proposed system in terms of energy saving with different spectral efficiencies, it can be seen that CFIM systems are superior to OFDM-IM and FIM in terms of energy saving as depicted in Fig \ref{fig:5.3.2}. Again, increasing the number of spreading codes $N_c$ improves significantly the energy efficiency. Furthermore, it can be seen that for $1$ bps/Hz, CFIM($4$,$4$) have $100\%$ energy saving. In fact, it means that one can transmit \emph{mapped bits} with no additional energy for modulated bits. Moreover, improving  spectral efficiency results in less energy saving, because the modulation order $M$ needs to be higher. As depicted in Fig. \ref{fig:5.3.3}, increasing the number of subcarriers does not achieve a better energy efficiency.

\begin{figure}[!t]
\centering
\includegraphics[width=3.6in,height=2.8in]{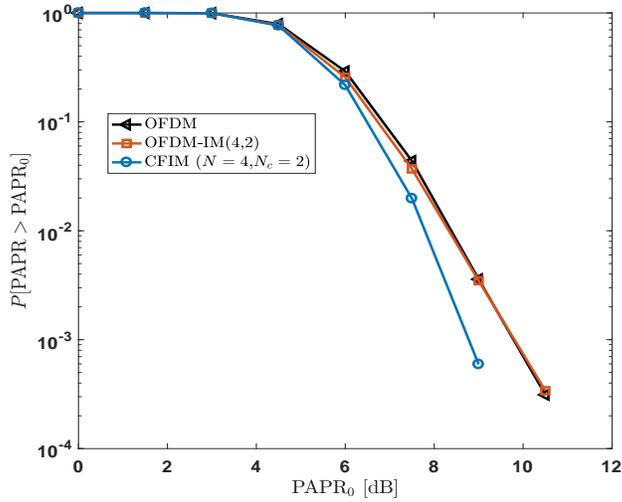}
\caption{CCDFs of the PAPR of the proposed CFIM($4$,$2$) in comparison with conventional OFDM and OFDM-IM($4$,$2$) using a modulation order of $M = 2$, number of blocks $K = 13$, the number of subcarriers per block is $N = 4$ and the FFT length is $N_T = 64$}
\label{fig:5.3.4}
\end{figure}

In Fig. \ref{fig:5.3.4}, we have evaluated the complementary cumulative distribution function (CCDF) of the PAPR in the proposed system in comparison with OFDM and OFDM-IM($4$,$2$) systems. It should be noted that these systems transmit $52$ bits per transmission and the FFT length is $N_T = 64$. It can be seen that the CFIM has the lowest PAPR reduction compared to these two systems. This is due to the fact that only one subcarrier per block is activated, which reduces the probability of having a high value of PAPR.  
 
\subsection{Performance of CFIM in Multiuser Communications}
Since spreading codes are used in the proposed system, it is easily extendible to the concept of multiuser communications. In synchronous transmission, each user transmits or receives the information in the same time slot. As such, orthogonal spreading codes may be implemented in order to mitigate multiuser interferences (MUI). Thus, assuming that each device uses CFIM systems, in both downlink and uplink transmission scenario, the BER performance is equal to equation \eqref{eq:3.1.1}, \textit{i.e.} this is equal to the performance of a single user CFIM transmission scenario. 

\begin{figure}[!htbp]
\centering
\includegraphics[width=3.6in,height=2.8in]{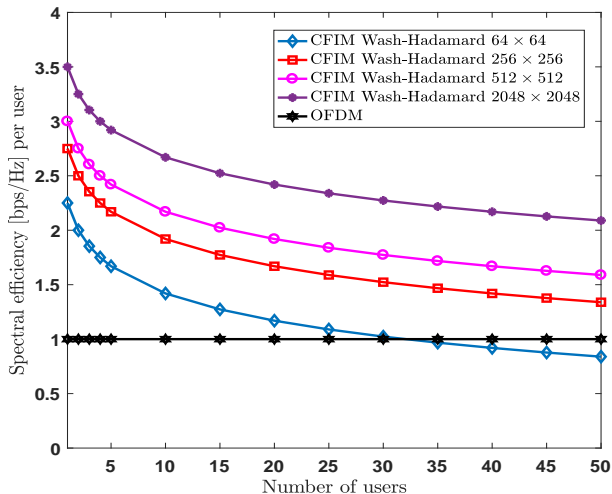}
\caption{Spectral efficiency in multiuser CFIM using a modulation order of $M = 2$ with a total of $N = 4$ subcarriers, and using Walsh-Hadamard matrices of various size. }
\label{fig:5.4.1}
\end{figure}

Furthermore, in a single-user scenario, we have seen that CFIM can enhance the spectral efficiency by increasing the number of spreading $N_c$. However, there is a trade-off between the number of users in the scenario and the spectral efficiency, because the number of orthogonal codes is a limited resource. Given the number of users and the size of a Walsh-Hadamard matrix, 
the maximal achievable spectral efficiency per user can be determined via equation \eqref{eq:SPE_EFF}. Fig. \ref{fig:5.4.1} depicts the spectral efficiency using a modulation order of $M = 2$ with a total of $N = 4$ subcarriers. It can be seen that CFIM system can significantly enhance the spectral efficiency compared to the conventional OFDM, with a large number of spreading codes available. It should be noted that the maximal achievable spectral efficiency decreases as the number of users increases. Notably, above $32$ users with a size of $64\times 64$ Walsh-Hadamard matrix, it is more efficient to use the conventional OFDM rather than CFIM systems to achieve higher spectral efficiency. 

\section{Conclusions}
\label{sect:VII}
In this paper, we have proposed a low-complexity index modulation based system that can significantly enhance the spectral and the energy efficiencies while maintaining reliability. This scheme is based on a joint index modulation: code and frequency index. The code index enhances the SE while the frequency index ensures a better EE by reducing high PAPR values. The proposed scheme suits IoT applications and can be extended to multiuser communications, as notably a remarkable performance has been shown in synchronous transmission. 

The introduced system may be built on OFDM platforms as this latter is abundant and easy to program. In short, the system is divided into $K$ blocks that operate $N$ subcarriers and $N_c$ spreading codes each, and use $M-$ary keying. At the transmitter side, each block contains an outgoing bit stream that is divided into three parts. The bits in the first part modulate an $M-$ary symbol that is actually transmitted through the channel. The bits in the second part are used to activate a subcarrier and the bits in the third part choose a spreading code. The $M-$ary symbol generated by the bits in the first part is first spread by the spreading code and then transmitted via the activated subcarrier. Note that a single subcarrier out of $N$ is activated per block and a spreading code out of $N_c$ is chosen. The remaining procedures are the same as those of classical OFDM. FFT is performed first at the receiver, then the signal is despread and square-law envelope detection is applied to estimate the code and frequency indices in order to recover the mapped bits, followed by a conventional $M-$ary demodulation process.

The acquired closed-form terms of the BER performance over fading channels is examined and confirmed by computer simulation. Moreover, analyses regarding complexity, SE and EE and complexity analyses have performed, where our findings indicate a drop in the peak to average power ratio. This PAPR cutback shows the suitability of the proposed approach to sensor-based IoT applications, where the portrait of both power and complexity should be preserved at small values. The modulation architecture presented in this work satisfies the requirements of 5G-based wireless systems as it minimizes PAPR and power consumption at the transmitter and demonstrates a satisfying overall performance in terms of reliability and high SE.

Some communication systems operate through asynchronous transmission. Under such a condition, performance severely degrades if orthogonal spreading codes such as Walsh-Hadamard or Zadoff-Chu sequences are used due to strong MUI. Therefore, using Gold codes will constitute an ideal alternative for such scenarios. Furthermore, most of the orthogonal multicarrier systems are sensitive to intercarrier interference caused by Doppler shift. Since CFIM systems are designed based on these two foundations, the overall performance may degrade. As a result, performance analysis and mitigation techniques for these cases will be considered in future works.


%

%

%
%

\ifCLASSOPTIONcaptionsoff
  \newpage
\fi



\bibliographystyle{IEEEtran}
\bibliography{Biblio_CFIM}

\begin{thebibliography}{10}
\providecommand{\url}[1]{#1}
\csname url@samestyle\endcsname
\providecommand{\newblock}{\relax}
\providecommand{\bibinfo}[2]{#2}
\providecommand{\BIBentrySTDinterwordspacing}{\spaceskip=0pt\relax}
\providecommand{\BIBentryALTinterwordstretchfactor}{4}
\providecommand{\BIBentryALTinterwordspacing}{\spaceskip=\fontdimen2\font plus
\BIBentryALTinterwordstretchfactor\fontdimen3\font minus
  \fontdimen4\font\relax}
\providecommand{\BIBforeignlanguage}[2]{{%
\expandafter\ifx\csname l@#1\endcsname\relax
\typeout{** WARNING: IEEEtran.bst: No hyphenation pattern has been}%
\typeout{** loaded for the language `#1'. Using the pattern for}%
\typeout{** the default language instead.}%
\else
\language=\csname l@#1\endcsname
\fi
#2}}
\providecommand{\BIBdecl}{\relax}
\BIBdecl

\bibitem{Rawat2014}
P.~Rawat, K.~D. Singh, H.~Chaouchi, and J.-M. Bonnin, ``Wireless sensor
  networks: a survey on recent developments and potential synergies,''
  \emph{The J. of Supercomput.}, vol.~68, pp. 1--48, 2014.

\bibitem{Tozlu2012}
S.~Tozlu, M.~Senel, W.~Mao, and A.~Keshavarzian, ``{W}i-{F}i enabled sensors
  for internet of things: A practical approach,'' \emph{IEEE Commun. Mag},
  vol.~50, pp. 134--143, 2012.

\bibitem{Chen2012}
Y.~Chen, ``Challenges and opportunities of internet of things,'' in \emph{17th
  Asia and south pacific Design Autom. Conf. (ASP-DAC)}, 2012.

\bibitem{Toh2001}
C.-K. Toh, ``Maximum battery life routing to support ubiquitous mobile
  computing in wireless ad hoc networks,'' \emph{IEEE Commun. Mag}, vol.~39,
  no.~6, pp. 138--147, 2001.

\bibitem{Basar2016}
E.~Basar, ``Index modulation techniques for 5{G} wireless networks,''
  \emph{IEEE Commun. Mag}, vol.~54, no.~7, pp. 168--175, 2016.

\bibitem{Mesleh2006}
R.~Mesleh, H.~Haas, C.~Ahn, and S.~Yun, ``Spatial modulation: a new low
  complexity spectral efficiency enhancing technique,'' in \emph{IEEE in. Conf.
  Commun. and Networking, ChinaCom'06}, 2006.

\bibitem{Jeyadeepan2008}
J.~Jeyadeepan, A.~Ghrayeb, and L.~Szczecinski, ``Spatial modulation: optimal
  detection and performance analysis,'' \emph{{IEEE} Commun. Lett.}, vol.~12,
  pp. 545--547, 2008.

\bibitem{Mesleh2008}
R.~Y. Mesleh, H.~Haas, S.~Sinamovic, C.~Ahn, and S.~Yun, ``Spatial
  modulation,'' \emph{{IEEE} Trans. Veh. Technol.}, vol. 574, pp. 2228--2241,
  2008.

\bibitem{Renzo2014}
M.~D. Renzo, H.~Haas, A.~Ghrayeb, S.~Sugiura, and L.~Hanzo, ``Spatial
  modulation for generalized {MIMO}: Challenges, opportunities, and
  implementation,'' \emph{Proc. IEEE}, vol. 102, no.~1, pp. 56--103, 2014.

\bibitem{Yang2015}
P.~Yang, M.~D. Renzo, Y.~Xiao, S.~Li, and L.~Hanzo, ``Design guidelines for
  spatial modulation,'' \emph{IEEE Communications Surveys and Tutorials},
  vol.~17, no.~1, pp. 6--26, 2015.

\bibitem{Abualhiga2009}
R.~Abualhiga and H.~Haas, ``Subcarrier-index modulation {OFDM},'' in
  \emph{Proc. IEEE PIMRC}, 2009, pp. 177--181.

\bibitem{Tsonev2011}
D.~Tsonev, S.~Sinanovic, and H.~Haas, ``Enhanced subcarrier index modulation
  ({SIM}) {OFDM},'' in \emph{Proc. IEEE GLOBECOM}, 2011, pp. 728--732.

\bibitem{Basar2013}
E.~Basar, U.~Aygolu, E.~Panayirci, and H.~V. Poor, ``Orthogonal frequency
  division multiplexing with index modulation,'' \emph{{IEEE} Trans. Signal
  Process.}, vol.~61, no.~22, pp. 5536--5549, 2013.

\bibitem{Kaddoum2016}
G.~Kaddoum, Y.~Nijsure, and H.~Tran, ``Generalized code index modulation
  technique for high-data-rate communication systems,'' \emph{{IEEE} Trans.
  Veh. Technol.}, vol.~65, no.~9, pp. 7000--7009, 2016.

\bibitem{Kaddoum2015a}
G.~Kaddoum, M.~Ahmed, and Y.~Nijsure, ``Code index modulation: a high data rate
  and energy efficient communication system,'' \emph{{IEEE} Commun. Lett.},
  vol.~19, no.~2, pp. 175--178, 2015.

\bibitem{Kaddoum2015}
G.~Kaddoum and E.~Soujeri, ``On the comparison between code-index modulation
  and spatial modulation techniques,'' in \emph{Proc. ICTRC}, 2015.

\bibitem{Soujeri2017}
E.~Soujeri, G.~Kaddoum, M.~Au, and M.~Hercerg, ``Frequency index modulation for
  low complexity low energy communication networks,'' \emph{IEEE Acess},
  vol.~PP, no.~99, pp. 1--11, 2017.

\bibitem{Boccardi2014}
F.~Boccardi, R.~H. Jr., A.~Lozano, T.~Marzetta, and P.~Popovski, ``Five
  disruptive technology directions for {5G},'' \emph{IEEE Commun. Mag},
  vol.~52, pp. 74--80, 2014.

\bibitem{Dai2015}
L.~Dai, B.~Wang, Y.~Yuan, S.~Han, C.-L. I, and Z.~Wang, ``Non-othogonal
  multiple acess for 5{G}: solutions, challenges, opportunities, and future
  research trends,'' \emph{IEEE Commun. Mag}, vol.~53, pp. 74--81, 2015.

\bibitem{Proakis2001}
J.~G. Proakis, \emph{Digital Communications}.\hskip 1em plus 0.5em minus
  0.4em\relax McGraw-Hill, 2001.

\bibitem{Simon2005}
M.~K. Simon and M.~Alouini, \emph{Digital Communication over Fading Channels:
  Second Edition}.\hskip 1em plus 0.5em minus 0.4em\relax John Wiley \& Sons,
  2005.

\bibitem{Lee1986}
P.~Lee, ``Computation of the bit error rate of coherent {M}-ary {PSK} with gray
  code bit mapping,'' \emph{{IEEE} Trans. Commun.}, vol.~34, no.~5, 1986.

\bibitem{Jiang2013}
T.~Jiang, C.~Li, and C.~Ni, ``Effect of {PAPR} reduction on spectram and energy
  efficiencies in {OFDM} systems with class-{A} {HPA} over {AWGN} channel,''
  \emph{IEEE Trans. Broadcast}, vol.~59, pp. 513--519, 2013.

\bibitem{Jiang2008}
T.~Jiang, M.~Guizani, H.~Chen, W.~Xiang, and Y.~Wu, ``Derivation of {PAPR}
  distribution for {OFDM} wireless systems based on extreme value theory,''
  \emph{IEEE Trans. Wireless Commun}, vol.~7, pp. 1298--1305, 2008.

\bibitem{Armstrong2002}
J.~Armstrong, ``Peak-to-average power reduction for {OFDM} by repeated clipping
  and frequency domain filtering,'' \emph{Electron. Lett.}, vol.~38, pp.
  246--247, 2002.

\end{thebibliography}
%

%
%
%
%
%




\end{document}